\documentclass[a4paper,11pt]{article}

\usepackage{jcappub} 
\usepackage[T1]{fontenc} 

\usepackage{amssymb}
\usepackage{graphics}
\usepackage{latexsym}
\usepackage{graphicx}
\usepackage{bm}
\usepackage{amsmath}
\usepackage{float}
\usepackage{graphicx}
\usepackage{setspace}
\usepackage{amsfonts}
\usepackage{fancyhdr}
\usepackage{layout}
\usepackage{epstopdf}
\usepackage{graphicx}
\usepackage{minitoc}
\usepackage{slashed}
\usepackage{mathrsfs}
\usepackage{eufrak}
\usepackage{mathtools}
\usepackage[hang,flushmargin]{footmisc} 
\usepackage[sans,nouppercase]{frontespizio}
\usepackage{extra}

\title{\boldmath Damping of cosmological tensor modes in Horndeski theories after GW170817}

\author{Mattia Scomparin}
\author{and Simone Vazzoler}

\affiliation{Studio Pointer s.r.l.,\\Via Tiziano Vecellio 13, Mogliano Veneto, Italy}

\emailAdd{mattia.scompa@gmail.com}
\emailAdd{svazzole@gmail.com}

\dedicated{Dedicated to Luciano}

\abstract{
This paper investigates the propagation of cosmological gravitational waves interacting with free-streaming neutrinos within the context of Horndeski theories of  gravity constrained by the detection of GW170817. 
We apply the theory of cosmological perturbations to explicitly derive the Einstein-Boltzmann equation for the damped propagation of first-order transverse traceless gravitational waves.
In contrast to general relativity, we argue that modified gravity can give rise to non-vanishing free-streaming damping effects during the cosmological matter dominated era.  
We also provide an analytic formula for the main multipole order with which modified gravity and free-streaming neutrinos damp the variety  of tensor correlation functions of the cosmic microwave background.
}
\begin{document}

\maketitle
\flushbottom

\section{Introduction}
\label{sec:intro}
Today's expansion of the universe \cite{Perlmutter:1998np, Riess:1998cb}, light element abundance \cite{Steigman:2000ha, Cyburt:2003fe, Coc:2017pxv} and temperature anisotropies on cosmic microwave background (CMB) \cite{Bartlett:1999vt,Lopez-Corredoira:2017rqn, Hu:1997mn, Santos:2017alg, Kamionkowski:1997vq, Liddle:1999qf, Hu:1996vr} are observational evidence accurately described in the context of the standard cosmological ($\Lambda$CDM) model of the hot Big Bang \cite{Scott:2005uf, Peebles:1998yv}. Within this model, the theory of general relativity (GR) is the standard theory of gravity  \cite{Weinberg:1972kfs}, and the existence of unknown  cosmological constant ($\Lambda$) and cold dark matter (CDM) energy components is required  to fit the growing amount of data provided by a wide range of cosmological probes \cite{Hinshaw:2012aka, Ade:2015xua, Ade:2015zua,Ishak:2018his}. 
Despite its  observational successes, the $\Lambda$CDM model exhibits deep theoretical difficulties involving, among others, questions such as the validity of GR on large scales of the universe \cite{Debono:2016vkp} and the consistence of the inflationary picture  \cite{Lake:2004xg} concerning the early origin of the cosmos \cite{Bull:2015stt}. 
These subjects have been largely studied over the past century, and several gravitational theories alternative to GR have been proposed \cite{Clifton:2011jh} as possible approach to address some of these outstanding  problems.

Scalar-tensor (ST) theories \cite{Clifton:2011jh} hold a prominent role with respect to these issues.
Such  modified gravity theories extend GR by introducing one or more additional scalar degrees of freedom \cite{Chauvineau:2015cha}, and their structure must be defined avoiding the presence of the so-called Ostrogradsky ghost instabilities \cite{Woodard:2015zca}.
Among the most sophisticated ghost-free ST frameworks developed for a single scalar degree of freedom, researchers  especially focus on the so-called Horndeski or ``generalized Galileon''  theories \cite{Horndeski:1974wa}.  These theories are the most general four-dimensional covariant ST theories of gravity yielding up to second-order equations of motion \cite{Deffayet:2013lga}, a sufficient requirement to avoid Ostrogradski ghost. It is worth noticing that such theories include GR, quintessence, k-essence, f(R) gravity, Brans-Dicke (BD) theories, and Galileons \cite{ Nunes:2017bwb} as special cases.

Generally, theories of modified gravity are characterized by different theoretical predictions for the propagation speed $c_{GW}$ of gravitational waves (GWs). In view of this evidence, the strong  constraint $-3\times10^{-15} \le c_{GW}/c -1 \le 7\times10^{-16}$, imposed by the recent multi-messenger detection of the gravitational GW170817 and electromagnetic GRB170817A signals emitted by neutron star mergers \cite{GBM:2017lvd, Monitor:2017mdv}, has ruled out all the theories predicting $c_{GW}$ which differs from the speed $c$ of light.  
More specifically, the impact of this bound on the Horndeski framework has been studied in refs. \cite{Ezquiaga:2017ekz, Creminelli:2017sry, Amendola:2017orw, Sakstein:2017xjx}, that found precise mathematical relations among the free functions of the Horndeski theories. In the following sections, we refer to such set of survived theories as the ``restricted'' Horndeski framework.

At the dawn of the new era of ``multi-messenger astronomy'' \cite{Fattoyev:2017jql,Piekarewicz:2018zbi}, scientists may investigate the cosmology of the early universe with innovative methods and experiments \cite{Allen:2018yvz} by making use of  newly combined information resulting from both electromagnetic and gravitational radiation.
In addition, the detection of primordial gravitational waves (PGWs) could provide fundamental information about the validity of numerous inflationary \cite{Linde:1984ir} and modified gravity models \cite{Joyce:2016vqv}. In this respect, the observation of the inflationary gravitational wave background (IGWB)  \cite{Hiramatsu:2006bd} could give a precise evidence of quantum gravity phenomenon \cite{Guzzetti:2016mkm}. 
Although PGWs are yet to be detected, it is expected that future generations of interferometers such as the Laser Interferometer Space Antenna (LISA) \cite{Audley:2017drz} and the Deci-hertz Interferometer Gravitational Wave Observatory (DECIGO) \cite{Kawamura:2011zz} could get closer to the experimental sensibility to detect them or place strong constraints on their amplitudes.

Against this backdrop, it is important that how PGWs travel through our universe be studied in view of modeling   physical phenomena that modify their dynamics and, consequently, the physical information they transport.
A non-negligible alteration is given by the damping effect sourced by the interaction of GWs  with cosmological free-streaming neutrinos \cite{Lancaster:2017ksf, Basboll:2008fx}, decoupled from electrons, positrons and photons at $T_{dec}\sim 2$ Mev. More specifically, first-order transverse traceless  terms in the neutrino anisotropic stress energy-momentum tensor modify the amplitude of  GWs, as shown in the context of GR by ref. \cite{Weinberg:2003ur}.
As a result, such interaction affects the main multipole order $\ell_k$ which appears in various correlation functions related to GWs signatures  on CMB \cite{Watanabe:2006qe, Pritchard:2004qp, Mangilli:2008bw}. 

Among other possible alterations, it has been verified that a wide range of modified gravity theories may predict significant changes on the propagation of cosmological GWs \cite{Saltas:2014dha,Pettorino:2014bka,Nunes:2018zot}. 
As many inflationary models have been built based on such theories, modified gravity effects must be considered in order to analyze the main mechanisms of PGWs production, propagation, and modification.

In this paper, we present a groundwork study on how the combined coexistence of these two types of sources, i.e.,  the modification of GR theory and the streaming of free cosmological neutrinos, damps the amplitude of PGWs.  We elaborate an extension of the results obtained by ref. \cite{Weinberg:2003ur} within the restricted Horndeski framework. We obtain a mathematical expression for the multipole order $\ell_k$  and derive the generalization of the Einstein-Boltzmann integro-differential equation for the propagation of damped cosmological GWs.  
Interestingly, we reach these results by maintaining total generality, without fixing a specific profile for the free functions of the restricted Horndeski theories.

The manuscript is organized as follows. In section \ref{sec:GW}, we discuss the restricted Horndeski framework and deduce the associated covariant scalar-tensor field equations.
Then,  in section \ref{Sec:FLRW}, we use the abovementioned results to study, up to the first order, the propagation of cosmological gravitational waves interacting with free-streaming neutrinos embedded in an FLRW cosmology. In section \ref{sec:damping}, we obtain the generalization of the Einstein-Boltzmann equation for the damped propagation of cosmological GWs.  Subsequently, a generalized analytic formula is deduced for the CMB multipole order $\ell_k$.
Finally, we close our analysis in the section \ref{sec:conc}.

In this work, we use the metric signature ($-$,$+$,$+$,$+$) and set the speed of light to unit. Greek indices run from 0 to 3, whereas Latin ones run from 1 to 3 and label spatial coordinates.


\section{Restricted Horndeski theories}\label{sec:GW}

The restricted Horndeski (rH) theories correspond to the most general Horndeski ST framework consistent with the tensor propagation speed $c_{GW}=1$ \cite{Ezquiaga:2017ekz, Creminelli:2017sry, Amendola:2017orw}. In this article, we focus on the action for rH theories that reads
\label{sec:fieldEquations}
\begin{equation}\label{eq:1}
S_{\mbox{\tiny rH}}\equiv\int d^4x \sqrt{-g} \left[\frac{1}{16\pi G}\sum_{i=2}^4\mathcal{L}_{\mbox{\tiny rH}}^i+\mathcal{L}_m\right]\,,
\end{equation}
where $g$ is the determinant of the metric tensor $g_{\mu\nu}$, $G$ is the gravitational constant, $\mathcal{L}_m$ is the matter Lagrangian density, and  
\begin{equation}\label{eq:20a}
\mathcal{L}_{\mbox{\tiny rH}}^2\equiv G_2(X,\vp)\qquad
\mathcal{L}_{\mbox{\tiny rH}}^3\equiv-\,G_3(X,\vp) \Box\vp\qquad
\mathcal{L}_{\mbox{\tiny rH}}^4\equiv G_4(\vp) R\,.
\end{equation}
We identify $R$ as the Ricci scalar, and the symbols $\nabla_\mu$ and $\Box\equiv \nabla_\mu\nabla^\mu$ stand for the covariant derivative and the d'Alembert operator, respectively. 

As shown in the above definition \eqref{eq:20a}, the three free functions $G_i$ depend on the scalar field $\vp$ and the canonical kinetic term $X\equiv -\tfrac{1}{2}\nabla_\mu\vp\nabla^\mu\vp$.

Clearly, by imposing the condition
\begin{equation}\label{eq:200}
G_2(X,\vp)=0\qquad G_3(X,\vp)=0\qquad G_{4}(\vp)=1\,,
\end{equation}
the well-known Einstein-Hilbert action for GR is recovered.

Oftentimes, the attempt to predict physical observables from general frameworks such as action \eqref{eq:1} requires setting case-specific profiles for the free functions and taking dedicated couplings of $\mathcal{L}_m$ with the other fields of the theory.
For simplification purposes, in this case we just assume $\mathcal{L}_m$ to be coupled with the metric tensor only.

\subsection{Covariant field equations}

Starting from  action \eqref{eq:1}, in this section, we obtain the explicit expression for the covariant field equations of motion.
Variation of action \eqref{eq:1} with respect to the scalar field $\delta S/\delta\vp=0$ leads to the scalar field equation \cite{Kobayashi:2011nu, McManus:2016kxu} 
\begin{equation}\label{eq:6}
\mathcal{J}=0\,,
\end{equation}
with 
\begin{equation}\label{eq:6h}
\mathcal{J}\equiv\sum_{i=2}^4\Big[\nabla^\mu j^{\,i}_\mu-p_\vp^{\,i}\Big]\,.
\end{equation}
In eq. \eqref{eq:6h} we identify the elementary currents $j_{\mu}^i$ and the elementary scalars $p_\vp^i$  as
\begin{eqnarray}
j^2_{\mu}&=& -\mathcal{L}_{2X} \nabla_{\mu}{\vp}\,,\\
j^3_{\mu} &=& -\mathcal{L}_{3X} \nabla_{\mu}{\vp}+G_{3X}\nabla_{\mu}{X}+2G_{3\vp} \nabla_{\mu}{\vp}\,,\\
j^4_{\mu} &=& 0\,,
\end{eqnarray}
and
\begin{eqnarray}
p_\vp^2 &=& G_{2\vp}\,,\\
p_\vp^3 &=& \nabla_{\mu}G_{3\vp} \nabla^{\mu}{\vp}\,,\\
p_\vp^4 &=& G_{4\vp}R\,,
\end{eqnarray}
with $G_{iX}\equiv\partial G_i/\partial X$, $G_{i,\vp}\equiv\partial G_i/\partial \vp$ and $\mathcal{L}_{i X}\equiv\partial \mathcal{L}_{i}/\partial X$. Note that in the case of a purely shift-symmetric theory, $G_4$ = const, as well as $G_{2,3}=G_{2,3}(X)$,  and therefore $p_\vp^i=0$.

Similarly, variation of action \eqref{eq:1} with respect to the metric tensor $\delta S/\delta g^{\mu\nu}=0$  leads to the tensor covariant field equation
\begin{equation}
\label{eq:7}
\mathcal{H}_{\mu\nu}=0\,,
\end{equation}
with
\begin{equation}
\label{eq:7h}
\mathcal{H}_{\mu\nu}\equiv\sum_{i=2}^4h_{\mu\nu}^{\,i}-8\pi G \,T_{\mu\nu}\,.
\end{equation}
In eq. \eqref{eq:7h}, $T_{\mu\nu}$ is the energy-momentum tensor defined as 
\begin{equation}\label{eq:emt}
T^{\mu\nu}\equiv\frac{2}{\sqrt{-g}}\frac{\delta(\sqrt{-g}\mathcal{L}_m)}{\delta g_{\mu\nu}}\,,
\end{equation}
and the three tensor densities $h_{\mu\nu}^{\,i}$ have expression
\begin{eqnarray}
h^2_{\mu \nu} &=&  - \tfrac{1}{2}G_{2X} \nabla_{\mu}{\vp} \nabla_{\nu}{\vp} - \tfrac{1}{2}G_2 g_{\mu \nu}\,,\\
h^3_{\mu \nu} &=& \tfrac{1}{2}G_{3X} \Box \vp \nabla_{\mu}{\vp} \nabla_{\nu}{\vp}+\tfrac{1}{2}\nabla_{\mu}{G_{3}} \nabla_{\nu}{\vp}+\tfrac{1}{2}\nabla_{\nu}{G_{3}} \nabla_{\mu}{\vp} - \tfrac{1}{2}g_{\mu \nu} \nabla_{\alpha}{G_{3}} \nabla^{\alpha}{\vp}\,,\\
h^4_{\mu \nu} &=& G_{4} G_{\mu \nu}+g_{\mu \nu} \left(G_{4\vp} \Box \vp-2X G_{4\vp\vp}\right)-G_{4\vp} \nabla_{\mu}{\nabla_{\nu}{\vp}}-G_{4\vp\vp} \nabla_{\mu}{\vp} \nabla_{\nu}{\vp}\,,
\end{eqnarray}
with $G_{\mu\nu}$ as the standard Einstein tensor.
Substituting the above expressions for  $j_{\mu}^i$, $p_\vp^i$, and $h_{\mu\nu}^{\,i}$ in eqs.  \eqref{eq:6} and \eqref{eq:7}, one can easily verify that they are defined up to second-order derivatives, so no ghost instabilities are propagated. 

As $\mathcal{L}_m$ is assumed to be coupled only with the metric, the energy-momentum tensor \eqref{eq:emt} satisfies the conservation law
\begin{eqnarray}\label{eq:13}
\mathcal{T}_\mu=0\,,
\end{eqnarray}
with $\mathcal{T}_\mu\equiv\nabla^{\nu}T_{\nu\mu}$.


\section{FLRW background and tensor perturbations}\label{Sec:FLRW}

In the context of the theory of cosmological perturbations, let us consider tensor perturbations around a spatially flat Friedmann Lema\^{i}tre Robertson Walker (FLRW) background spacetime.
Generally, these perturbations can be described up to the first order by the perturbed non-vanishing metric components 
\begin{equation}\label{eq:10a}
g_{00}=g^{(0)}_{00}\qquad g_{ij}=g^{(0)}_{ij}+g^{(1)}_{ij}\,,
\end{equation}
with
\begin{equation}\label{eq:10}
g_{00}^{(0)}=-1\qquad
g_{ij}^{(0)}=a^2\delta_{ij}\qquad
g_{ij}^{(1)}=a^2h_{ij}\,.
\end{equation}
As usual, $a\equiv a(t)$ is the scale factor, $\delta_{ij}$ is the Kronecker delta function, and $h_{ij}\equiv h_{ij}(\vec{x},t)$ is treated as a small perturbation $|h_{ij}|\ll1$. 
We assume that the scalar field $\vp$ contributes only to background dynamics and therefore its profile is purely isotropic:
\begin{eqnarray}\label{eq:19}
\vp=\psi(t)\,.
\end{eqnarray}
Hereafter, we refer to background and first-order perturbed quantities using, respectively, the superscripts $^{(0)}\sim\mathcal{O}(1)$ and $^{(1)}\sim\mathcal{O}(h)$.

For the purposes of this analysis, we also consider the perturbations of the energy-momentum tensor $T_{\mu\nu}$ around a cosmological perfect fluid background. Up to the first order, the only non-vanishing components are 
\begin{equation}\label{eq:11a}
T_{00}=T^{(0)}_{00}\qquad T_{ij}=T^{(0)}_{ij}+T^{(1)}_{ij}\,,
\end{equation}
with
\begin{equation}
\label{eq:11}
T_{00}^{(0)}=\rho\qquad
T_{ij}^{(0)}=a^2P\delta_{ij}\qquad
T_{ij}^{(1)}=a^2\big[Ph_{ij}+\pi_{ij}\big]\,,
\end{equation}
where $\rho=\rho(t)$ is the background energy density and $P=P(t)$  is the pressure.
It is worth noticing that in eq. \eqref{eq:11},  we identify with $\pi_{ij}\equiv\pi_{ij}(\vec{x},t)$  all the first-order anisotropic contributions to the energy-momentum tensor. For a perfect fluid $\pi_{ij}=0$.

Using eqs. \eqref{eq:10a}, \eqref{eq:19}, and \eqref{eq:11a}, we can perturb the field equations \eqref{eq:6}, \eqref{eq:7} and \eqref{eq:13} up to the first order as
\begin{equation}\label{eq34}
\mathcal{J}^{(0)}+\mathcal{J}^{(1)}=0\qquad
\mathcal{H}_{\mu\nu}^{(0)}+\mathcal{H}_{\mu\nu}^{(1)}=0\qquad 
\mathcal{T}_{\mu}^{(0)}+\mathcal{T}_{\mu}^{(1)}=0\,.
\end{equation}

In this article we work in the so-called transverse traceless (TT) gauge, considering only tensor perturbations such that 
\begin{equation}\label{eq:gauge}
h^i\!\,_i=0\qquad \partial^ih_{ij}=0\qquad \pi^i\!\,_i=0\qquad \partial^i\pi_{ij}=0\,,
\end{equation}
with $h^i\!\,_i=\delta^{ij}h_{ij}$ and $\pi^i\!\,_i=\delta^{ij}\pi_{ij}$ null traces. 

By combining each perturbative order of eqs. \eqref{eq:10}, \eqref{eq:11}, and  \eqref{eq:19} with eq. \eqref{eq34}, we obtain the non-trivial background cosmological equations
\begin{equation}\label{eq:orderZero}
\mathcal{J}^{(0)}=0\qquad \mathcal{H}^{(0)}_{00}=0\qquad \mathcal{H}^{(0)}_{ij}=0\qquad \mathcal{T}^{(0)}_{0}=0\,,
\end{equation}
and the non-trivial  first-order TT equation
\begin{equation}\label{eq:orderOne}
\mathcal{H}^{(1)}_{ij}=0\,.
\end{equation}


\subsection{Background cosmology}

By referring to expression \eqref{eq:orderZero}, the set of functionally independent cosmological equations is composed of a conservation law and two modified Fridmann equations
\begin{eqnarray}
\label{eq:29}
\dot{\rho}(t)&=&-3H(t)\big[\rho(t)+P(t)\big]\,,\\
\label{eq:30}
H^2(t)+\lambda_1(t)H(t)+\lambda_2(t)&=&\tfrac{8}{3}\pi\,\mathcal{G}(t)\rho(t)\,,\\
\label{eq:31}
\dot{H}(t)+H^2(t)+\lambda_3(t)H(t)+\lambda_4(t)&=&-\tfrac{4}{3}\pi\,\mathcal{G}(t)\big[\rho(t)+3P(t)\big]\,,
\end{eqnarray}
where $H\equiv\dot{a}/a$ is the Hubble expansion rate and $\mathcal{G}\equiv G/G_{4}\!^{(0)}$ is the running gravitational constant.\footnote{Eq. \eqref{eq:29} correspond to the conservation law $\mathcal{T}_0^{(0)}=0$, and eqs. \eqref{eq:30} and \eqref{eq:31} can be identified as the modified Fridmann equations $\mathcal{H}_{00}^{(0)}=0$ and $-\frac{1}{2a^2}\mathcal{H}_{ij}^{(0)}-\frac{1}{6}\mathcal{H}_{00}^{(0)}=0$. Background components \eqref{eq:orderZero} are not functionally independent because $\dot{\mathcal{H}}_{00}^{(0)}+3\,\dot{a}a^{-1}(\mathcal{H}_{00}^{(0)}+\mathcal{H}_{ij}^{(0)}a^{-2})+\frac{1}{2}\dot\psi\mathcal{J}^{(0)}-8\pi G \mathcal{T}_0^{(0)}=0$.}   Dot denotes the derivative with respect to the time coordinate. The explicit mathematical expressions for cosmological coefficients $\lambda_i$ are given by
\begin{eqnarray}
\label{eq:36a}
\lambda_1&\equiv&\left[G_{4\vp}^{(0)} \dot{\psi}-\tfrac{1}{2}G_{3X}^{(0)} {\dot{\psi}}^{3}\right]G_4^{(0)\,-1}\,,\\
\label{eq:36b}
\lambda_2&\equiv&\left[\tfrac{1}{6}G_2^{(0)}- \tfrac{1}{6}G_{2X}^{(0)} {\dot{\psi}}^{2}+\tfrac{1}{6}G_{3\vp}^{(0)} {\dot{\psi}}^{2}\right]G_4^{(0)\,-1}\,,\\
\label{eq:36c}
\lambda_3&\equiv&\left[\tfrac{1}{2}G_{4\vp}^{(0)} \dot{\psi}+\tfrac{1}{4}G_{3X}^{(0)} {\dot{\psi}}^{3}\right]G_4^{(0)\,-1}\,,\\
\label{eq:36d}
\lambda_4&\equiv&\left[\tfrac{1}{6}G_2^{(0)} - \tfrac{1}{4}G_{3X}^{(0)} \ddot{\psi} {\dot{\psi}}^{2} - \tfrac{1}{3}G_{3\vp}^{(0)} {\dot{\psi}}^{2}+\tfrac{1}{2}G_{4\vp}^{(0)} \ddot{\psi}+\tfrac{1}{2}G_{4\vp\vp}^{(0)} {\dot{\psi}}^{2}\right]G_4^{(0)\,-1}\,.
\end{eqnarray}
with $G_{i}\!^{(0)}= G_i(\psi,X^{(0)})$, which depend only on the time coordinate through their dependence on $\psi(t)$ and $X^{(0)}(t)=\tfrac{1}{2}\dot{\psi}^2(t)$.
To avoid divergences on the right-hand sides of the above definitions, we require that $G_4\!^{(0)} \neq 0$.
Relation \eqref{eq:200} also shows that in the GR limit, the standard Fridmann equations \cite{Weinberg:1972kfs} are recovered because $\mathcal{G}=G$ and $\lambda_i=0$.

Considering that we have three eqs. \eqref{eq:29}, \eqref{eq:30}, and \eqref{eq:31}, for the four unknown functions $(a,P,\rho,\psi)$, we need an additional equation of state (EOS). Decomposing $P\equiv\sum_{i}P_i$ and $\rho\equiv\sum_{i}\rho_i$ as the sum of separate sources, we assume the so-called barotropic EOS:
 \begin{equation}\label{eq:33}
P_i(t)\equiv w_i \,\rho_i(t)\,,
\end{equation}
where $w_i$  is a constant so that $w_m=0$ leads to a vanishing pressure, $P_m$, corresponding to the matter dominated ($m$) model, and $w_\gamma=w_\nu=1/3$, the radiation-dominated model. By referring to radiation, we mean photons ($\gamma$) and neutrinos ($\nu$).

From eqs. \eqref{eq:29} and \eqref{eq:33}, it follows that
\begin{equation}\label{eq:52}
\rho_i(t)=\rho_{0,i}\left[\frac{a_0}{a(t)}\right]^{3(1+w_i)}\,,
\end{equation}
with $a_0\equiv a(t_0)$ and $\rho_{0,i}\equiv\rho_{i}(t_0)$. Hereafter, the subscript zero denotes quantities evaluated at present day epoch $t_0$.

After a little algebra with eq. \eqref{eq:52}, we put eq. \eqref{eq:30} in the form
\begin{equation}\label{eq:51}
H^2(t)+\lambda_1(t)H(t)+\lambda_2(t)=E^2(t)\,\Pi_0H_0^2\,,
\end{equation}
where we identify
\begin{equation}\label{eq:104}
E^2(t)\equiv\left[\frac{a_0}{a(t)}\right]^3\frac{\Omega_{0m}}{\Omega_0}+\left[\frac{a_0}{a(t)}\right]^4\frac{\Omega_{0\gamma}+\Omega_{0\nu}}{\Omega_0}\,,
\end{equation}
and
\begin{equation}
\label{eq:130}
\Pi(t)\equiv1+ \frac{\lambda_1(t)}{H(t)}+ \frac{\lambda_2(t)}{H^2(t)}\,,
\end{equation}
with $\Omega_i\equiv\rho_i/\varrho_c$ being the density parameters defined in terms of the critical density $\varrho_c\equiv3H^2/(8\pi G)$.
Generally, one can note that the density parameters $\Omega_i$ can be related to their respective  modified density parameters $\omega_i\equiv \rho_i/\rho_c$  through the mathematical relation
\begin{equation}\label{eq:90}
\omega_i(t)=\frac{1}{ \Pi(t)\,G_4^{(0)}(t)}\,\Omega_i(t)\,,
\end{equation}
where $\rho_c\equiv  \Pi\,G_4\!^{(0)}\varrho_c$ is the modified critical density, so that  eq. \eqref{eq:30} can be alternatively rewritten as 
\begin{equation}\label{eq:112}
\sum_{i}\frac{\omega_{i}(t)}{\omega(t)}=\sum_{i}\frac{\Omega_{i}(t)}{\Omega(t)}=1\,.
\end{equation}
Here, $\omega\equiv\sum_i\omega_i$ and $\Omega\equiv\sum_i\Omega_i$. One can observe therefore that in the GR limit \eqref{eq:200}, we have  $\Pi=1$, and hence $\rho_{c}=\varrho_c$ as well as $\omega_{i}=\Omega_i$.


\subsection{Tensor modes and first-order free-streaming neutrinos anisotropic inertia}

In the presence of TT terms in the anisotropic stress tensor, eq. \eqref{eq:orderOne} leads to
\begin{equation}\label{eq:20}
\ddot{h}_{ij}(\vec{x},t)+\left[3\,\frac{\dot{a}(t)}{a(t)}+\kappa(t)\right]\dot{h}_{ij}(\vec{x},t)-\frac{\nabla^2h_{ij}(\vec{x},t)}{a^2(t)}=16\pi\,\mathcal{G}(t)\,\pi_{ij}(\vec{x},t)\,,
\end{equation}
where $\nabla^2\equiv\p_k\p^k$ is the Laplacian operator and $\kappa\equiv G_{4\vp}\!^{(0)}\dot{\psi}/G_4\!^{(0)}$ is a shear viscosity parameter introduced by the rH framework.\footnote{Eq. \eqref{eq:20} corresponds to $2 a^{-2}(\mathcal{H}^{(1)}_{ij}-\tfrac{1}{3}\mathcal{H}^{(0)}_{mn}\delta^{mn}h_{ij}) =0$} Clearly, $\kappa=0$ in the GR limit \eqref{eq:200}.

To proceed further, we consider the Fourier expansions
\begin{equation}\label{eq:22}
h_{ij}(\vec{k},t)\equiv\int\frac{d^3k}{(2\pi)^3}e^{\,i\vec{k}\cdot\vec{x}}h_{ij}(\vec{x},t)\qquad
\pi_{ij}(\vec{k},t)\equiv\int\frac{d^3k}{(2\pi)^3}e^{\,i\vec{k}\cdot\vec{x}}\pi_{ij}(\vec{x},t)\,,
\end{equation}
and the new time coordinate 
\begin{equation}\label{eq:37}
u(t)\equiv k\eta(t),
\end{equation}
where $\vec{k}$ is a co-moving wave number with  $k\equiv |\vec{k}|$, $\eta(t)\equiv \int\!_{t_*}^t d\tau/a(\tau)$ is the conformal time, and $t_*$ is an arbitrary time scale. Thus, in the Fourier space  $(u,\vec{k})$ eq. \eqref{eq:20} becomes 
 \begin{equation}\label{eq:23}
\frac{d^2h_{ij}(u,\vec{k})}{du^2}+\left[\frac{2}{a(u)}\frac{da(u)}{du}+\kappa(u)\right]\frac{dh_{ij}(u,\vec{k})}{du}+h_{ij}(u,\vec{k})=16\pi\mathcal{G}(u)\frac{a^2(u)}{k^2}\pi_{ij}(u,\vec{k})\,.
\end{equation}
From expression \eqref{eq:23}, it is clear that, in the context of rH theories, interactions of GWs  with matter or radiation fields are possible only with a non-vanishing anisotropic contribution $\pi_{ij}$ to the energy-momentum tensor. 

In this respect, cosmological  free-streaming neutrinos, decoupled  with positrons, electrons, and photons at the time $u_{dec}\sim0$ ($ T_{dec}\sim 2$Mev), are one possible source of such anisotropic contribution.
As shown by refs. \cite{Weinberg:2003ur, Watanabe:2006qe, Pritchard:2004qp}, when treating relativistic neutrino gas by classical kinetic theory, the linearized collisionless Boltzmann equation gives 
\begin{equation}\label{eq:38}
\pi_{ij}(u)=-4\rho_\nu(u)\int_0^u \mathcal{K}\big[U-u\big] \,\frac{dh_{ij}(U)}{dU}dU\,,
\end{equation}
where $\rho_\nu$ indicates the neutrino energy density and $\mathcal{K}$ is the kernel
\begin{equation}\label{eq:39}
\mathcal{K}\big[s\big]\equiv-\frac{\sin s}{s^3}-3\frac{\cos s}{s^4}+3\frac{\sin s}{s^5}\,.
\end{equation}

\section{Damping of tensor modes caused by free-streaming neutrinos}
\label{sec:damping}

Assuming that at the initial time $u_{dec}$ all modes of cosmological interest are out of the cosmological horizon; i.e., $ k\ll aH$, and following the same procedure described by ref. \cite{Watanabe:2006qe}, one finds that if $1/(a^2\,G\!^{(0)}_4)$ behaves like a decay mode that rapidly approaches zero, then outside the horizon, $h_{ij}$ remains constant. 
Therefore, taking on this hypothesis; i.e., after the end of inflation, the amplitude of tensor fluctuations is conserved until the mode re-enters the horizon, one can conclude that \cite{Weinberg:2003ur}
\begin{equation}\label{eq:301}
h_{ij}(u)\equiv h_{ij}(0)\chi(u)\,,
\end{equation}
which computed inside eq. \eqref{eq:23} gives the Einstein-Boltzmann equation:
\begin{equation}\label{eq:24}
\chi''(u)+\left[2\frac{a'(u)}{a(u)}+\kappa(u)\right]\chi'(u)+\chi(u)=-64\pi\mathcal{G}(u)\frac{a^2(u)}{k^2}\rho_\nu(u)\int_0^u \mathcal{K}\big[U-u\big] \,\chi'(U)\,dU\,,
\end{equation}
with $\chi(0)=\chi'(0)=0$. For simplification purposes, from now on we use primes to indicate derivatives with respect to arguments we are focusing on.
As eq. \eqref{eq:24} is balanced by $\rho_\nu$, it is useful to introduce the neutrino distribution density $f_\nu(u)\equiv\rho_\nu(u)/\rho(u)$, which inserted into the modified Fridmann eq. \eqref{eq:30} yields
\begin{equation}\label{eq:49}
\rho_\nu(u)=\frac{3\, k^2}{8\pi\,a^2(u)\,\mathcal{G}(u)}f_\nu(u)\Upsilon_k(u)\,,
\end{equation}
with 
\begin{equation}
\Upsilon_k(u)\equiv \lambda_1(u)\frac{a'(u)}{k}+\lambda_2(u)\frac{a^2(u)}{k^2}+\frac{a'^{\,2}(u)}{a^2(u)}\,.
\end{equation}
From eqs. \eqref{eq:49} and \eqref{eq:24}, it follows that
 \begin{eqnarray}\label{eq:50}
\chi''(u)&+&\left[2\frac{a'(u)}{a(u)}+\kappa(u)\right]\chi'(u)+\chi(u)=-24f_\nu(u)\Upsilon_k(u)\int_0^u \mathcal{K}\big[U-u\big] \,\chi'(U)\,dU\,,
\end{eqnarray}
where the dependence on the running gravitational constant $\mathcal{G}$ disappears. 
The GR limit of our eq. \eqref{eq:50} has been used by the authors of refs. \cite{Weinberg:2003ur,Watanabe:2006qe} to study the damping of GWs whose wavelengths are short enough to re-enter the cosmological horizon during the radiation dominated era.

To move on and deal with GWs that enter the cosmological horizon during the matter dominated era, let us introduce the new scalar variable
\begin{equation}\label{eq:80}
y(u)\equiv\frac{a(u)}{a_{eq}}\,,
\end{equation}
where we have identified $a_{eq}\equiv a(t_{eq})$ with $t_{eq}$ as the matter-radiation equivalence time such that $\rho_{m}(t_{eq})=\rho_{\gamma}(t_{eq})+\rho_{\nu}(t_{eq})$.
The explicit mathematical relation between $y$ and the $u$ variable can be obtained by solving the modified Fridmann eq. \eqref{eq:30}, which appropriately worked, can be expressed as follows
\begin{equation}\label{eq:57}
y''(u)^2+\mathcal{A}_k\,\lambda_1(u)\,y^2(u)\,y'(u)+\mathcal{A}_k^2\,\lambda_2(u)\,y^4(u)=\frac{y(u)+1}{\mathcal{Q}^2_k}\,.
\end{equation}
The time-independent coefficients $\mathcal{A}_k$ and $\mathcal{Q}^2_k$ are
\begin{equation}\label{eq:70}
\mathcal{A}_k\equiv\frac{a_{0}}{k(1+z_{eq})}\qquad\mathcal{Q}^2_k\equiv\frac{k^2\,\Omega_0}{a_0^2\,H_0^2\,\Pi_0\,\Omega_{0m}\left(1+z_{eq}\right)}\,,
\end{equation}
with $z_{eq}\equiv z(t_{eq})$ the matter-radiation equivalence redshift such that $1+z_{eq}=a_0/a_{eq}$.

After some algebraic manipulations, eq. \eqref{eq:57} can be simplified to read
\begin{equation}\label{eq:59}
u'_\pm(y)=\frac{\mathcal{Q}_k}{\sqrt{y+1}}\left[-\mathcal{Z}\frac{\lambda_1(y)\,y^2}{\sqrt{y+1}}\pm\sqrt{1+\mathcal{Z}^2\frac{\big[\lambda_1^2(y)-4\lambda_2(y)\big]\,y^4}{y+1}}\,\right]^{-1}\,,
\end{equation}
with 
\begin{equation}\label{eq:zzz}
\mathcal{Z}\equiv\frac{1}{2}\mathcal{A}_k\mathcal{Q}_k
\end{equation}
 which, as shown in the above definition \eqref{eq:70}, remains  constant over time and is mostly scale-invariant as k-independent.

Being the solution \eqref{eq:59} formally described by two branches, we can focus on the positive sign $du_+/dy$ that is consistent with the sign of the GR solution, as obtained in ref. \cite{Weinberg:2003ur}. 
Finally, eq. \eqref{eq:50} becomes
\begin{equation}\label{eq:75}
(y+1)\chi''(y)+\frac{\delta_{1}(4+5y)}{2y}\chi'(y)+\delta_2\mathcal{Q}_k^2\,\chi(y)=-24 \frac{\delta_3f_{0,\nu}}{y^2}\int_0^y \mathcal{K}\big[U(Y)-u(y)\big] \chi'(Y)\,dY\!,
\end{equation}
with $\chi(y)=\chi[u(y)]$ the amplitude. The initial conditions are $\chi(0)=0$ and $\chi'(0)=0$. In eq. \eqref{eq:75},  we point out the presence of the three rH parameters $\delta_{j}=\delta_{j}(y)$, whose explicit expression is 
\begin{eqnarray}
\label{eq:d1}
\delta_1&\equiv&\Big\{\mathcal{Z}^{2} y^{4} (3 y + 4) \big[4 \lambda_2(y) - \lambda_1^{2}(y)\big] + 2 y (y + 1) \big\{-\mathcal{Q}_k\kappa(y)\,\sigma(y)+\mathcal{Z}^2y^4\big[2\lambda_2(y)\nonumber\\
& &-\lambda_1(y)\lambda_1'(y)\big]+\mathcal{Z}y\,\sigma(y)\big[y\lambda_1'(y)+4\lambda_1(y)\big]\big\} - (5 y + 4)\, \sigma^{2}(y)\Big\}\nonumber\\
& &\Big\{\sigma(y)\,(5 y + 4)\big[\mathcal{Z} y^{2} \lambda_1(y) - \sigma(y)\big]\Big\}^{-1}\!,\\
\label{eq:d2}
\delta_2&\equiv&(y + 1)\Big\{\mathcal{Z} y^{2} \lambda_1(y) - \sigma(y)\Big\}^{-2}\!,\\
\delta_3&\equiv& \Big\{4\mathcal{Z}^2\, y^{4} \lambda_2(y) - 2\mathcal{Z}\, y^{2} \big[\mathcal{Z} y^{2} \lambda_1(y) - \sigma(y)\big] \lambda_1(y) +  \big[\mathcal{Z} y^{2} \lambda_1(y) - \sigma(y)\big]^{2}\Big\}\nonumber\\
\label{eq:d3}
& &\Big\{\mathcal{Z} y^{2} \lambda_1(y) - \sigma(y)\Big\}^{-2}\!,
\end{eqnarray}
with  
\begin{equation}
\sigma(y)\equiv\sqrt{y + 1- \mathcal{Z}^{2}\, y^{4} \big[4 \lambda_2(y) - \lambda_1^{2}(y)\big] }\,.
\end{equation}

The GR limit \eqref{eq:200} of eqs. \eqref{eq:d1},  \eqref{eq:d2},  and  \eqref{eq:d3} leads to  $\delta_j=1$, and hence, we can exactly recover the Einstein-Boltzmann equation obtained by ref.  \cite{Weinberg:2003ur}. 
When compared to such limit, eq. \eqref{eq:75} exhibits important features that are worth examining.

First,  the rH framework presents in eq. \eqref{eq:75} the same  integro-differential structure, with respect to the amplitude $\chi$, of the analogous result obtained for GR. 
Therefore, one now needs facing the presence of the $\delta_j$ parameters, which make the attempt to computationally solve eq. \eqref{eq:75}, without loss of generality, problematic.
Indeed, up to now, such parameters are a priori completely free as their structure is  related to the rH free functions $G_i$ through $\kappa$ and $\lambda_i$.

Second, though for GR the damping effects are unimportant during the matter dominated era as $\delta_3=1$ and $f_\nu\sim f_{0,\nu}/y\rightarrow0$, our general result  shows that $\delta_3$ can suppress such decay, giving rise to non-vanishing effects in such era. 
One can consider this behavior as  an important phenomenological signature that distinguishes between the GR action and modified gravity theories.

 Finally, we observe that, in contrast to GR, eq. \eqref{eq:75} relates $\chi$ and $\mathcal{Q}_k$ by explicitly involving cosmological parameters as the rH framework introduces the additional cosmological coefficient $\mathcal{Z}$.

In the next section, we briefly analyze the connection between the damping effects described by eq.  \eqref{eq:75} and the various correlation functions that parameterize GWs signatures on CMB in the context of the rH framework. In this regard, according to what was discussed in refs. \cite{Weinberg:2003ur,Watanabe:2006qe}, we deem relevant to asses which CMB multipole orders $\ell_k$ are mainly dominated by such damping effects.

\subsection{Multipole order effects} \label{sec:multipole}

In this section we provide an analytic formula for $\ell_k$, the CMB multipole order mainly involved through the tensor damping effects of eq. \eqref{eq:75} during the matter-radiation last-scattering epoch $t_{ls}$.

Starting from eq. \eqref{eq:51} and as directly resulting from its evaluation at the equivalence redshift $z_{eq}$, it follows that
\begin{equation}\label{eq:120}
H^2(z)+\lambda_1(z)H(z)+\lambda_2(z)=\frac{H^2_{eq}\Pi_{eq}}{E^2_{eq}}E^2(z)\,.
\end{equation}
We point out that $E^2_{eq}$ can be rewritten in a more direct fashion than the trivial evaluation of eq. \eqref{eq:104} at $t_{eq}$. The easier way to proceed is by applying the mathematical relation $\omega_{eq,m}=\omega_{eq,\gamma}+\omega_{eq,\nu}$, to get
\begin{equation}\label{eq:eeq}
E^2_{eq}=2\big(1+z_{eq}\big)^3\frac{\Omega_{0m}}{\Omega_0}\,.
\end{equation}
To now derive an explicit expression for $H$,  one needs solving eq. \eqref{eq:120} for $H$. The result is
\begin{equation}\label{eq:140}
H_\pm(z)=\sqrt{\Pi_{eq}}\,\frac{H_{eq}}{E_{eq}}\,\Sigma_\pm(z)\,,
\end{equation}
with
\begin{equation}\label{eq:134}
\Sigma_\pm(z)\equiv-\frac{E_{eq}}{2\sqrt{\Pi_{eq}}H_{eq}}\,\lambda_1(z)\pm\sqrt{E^2(z)+\frac{E_{eq}^2}{4H_{eq}^2}\Big[\lambda_1^2(z)-4\lambda_2(z)\Big]}\,.
\end{equation}
Without loss of generality, we consider the positive branch $H_+$ as consistent with the sign of the GR solution obtained in ref. \cite{Weinberg:2003ur}. 

Subsequently, we evaluate expression \eqref{eq:120} at $t_0$, and as $E^2_{0}=1$, and obtain
\begin{equation}\label{eq:131}
\Pi_0=\frac{1}{E_{eq}^2}\left(\frac{H_{eq}}{H_{0}}\right)^2\Pi_{eq}\,.
\end{equation}
Hence, by computing eq. \eqref{eq:131} inside the definition of $\mathcal{Q}_k$  proposed in eq. \eqref{eq:70}, we obtain 
\begin{equation}\label{eq:137}
\frac{k}{k_{eq}}=\sqrt{\frac{\Pi_{eq}}{2}}\,\mathcal{Q}_k\,,
\end{equation}
where we identify $k_{eq}\equiv a_{eq}H_{eq}$.

The main multipole order $\ell_k$ that receives a dominant contribution at the last-scattering epoch $t_{ls}$ can be defined as
\begin{equation}\label{eq:136}
\ell_k\equiv\frac{d_{A,ls}}{a_{ls}}\,k\,,
\end{equation}
with  $a_{ls}\equiv a(t_{ls})$ and $d_{A,ls}=d_{A}(z_{ls})$. In our notation, $d_{A}(z)$ corresponds to the so-called angular distance that, as function of the redshift, is commonly defined as \cite{Weinberg:1972kfs}
\begin{equation}\label{eq:139}
d_{A}(z)\equiv \frac{1}{1+z}\int_0^z\frac{d\bar{z}}{H(\bar{z})}\,.
\end{equation}
For the matter-radiation equivalence epoch,  $\ell_{eq}\equiv \ell_{k_{eq}}$, eqs. \eqref{eq:137} and \eqref{eq:136} give
\begin{equation}\label{eq:138}
\ell_k=\sqrt{\frac{\Pi_{eq}}{2}}\,\mathcal{Q}_k\,\ell_{eq}\,.
\end{equation}
Therefore, from eqs. \eqref{eq:140} and  \eqref{eq:139}, one may write a general expression for $\ell_{eq}$  as
\begin{equation}
\ell_{eq}=\frac{E_{eq}}{\sqrt{\Pi_{eq}}\big(1+z_{eq}\big)}\int_0^{z_{ls}}\frac{d\bar{z}}{\Sigma_+(\bar{z})}\,,
\end{equation}
which, computed inside eq. \eqref{eq:139}, finally leads to 
\begin{equation}
\ell_k=\vartheta\,\mathcal{Q}_{k}\,.
\end{equation}
In this result, we introduce the constant factor
\begin{equation}\label{eq:mm}
\vartheta \equiv \sqrt{2}\,\frac{H_{eq}\,E_{eq}}{1+z_{eq}}\int_0^{z_{ls}}d\bar{z}\left\{-\frac{E_{eq}}{\sqrt{\Pi_{eq}}}\,\lambda_1(\bar{z})+\sqrt{4\,H_{eq}^2\,E^2(\bar{z})+E_{eq}^2\big[\lambda_1^2(\bar{z})-4\lambda_2(\bar{z})\big]}\right\}^{-1}\!,
\end{equation}
with $E_{eq}$ defined in eq. \eqref{eq:eeq} and $H_{eq}$ as well  as $\Pi_{eq}$ deductible from eqs. \eqref{eq:51} and \eqref{eq:130}, respectively, evaluated at $t_{eq}$. 

Against this framework, it is clear that $\vartheta$ can be seen as a conversion factor from $\ell_k$ to $\mathcal{Q}_{k}$, which depends on cosmological parameters. The benefit of using such result lies on a direct mathematical relation between the solutions of eq. \eqref{eq:75} and the various values of $\ell_k$ on which they depend through $\mathcal{Q}_{k}$. Interestingly, as in the case of the GR result of ref \cite{Weinberg:2003ur}, the rH framework preserves a linear dependence of $\ell_k$ on only $\mathcal{Q}_{k}$. In \eqref{eq:mm} the $\mathcal{Z}$ parameter does not provide any contribution.

\section{Discussion and Conclusions}
\label{sec:conc}

In this paper we studied the propagation of cosmological GWs interacting with free-streaming neutrinos within the framework of rH theories of gravity. Firstly, we deduced the covariant scalar-tensor field equations of such theories; subsequently to ref. \cite{Weinberg:2003ur}, we analyzed the propagation of first-order tensor perturbations embedded in a spatially flat FLRW cosmology.  We also assumed the presence of first-order TT terms in the anisotropic stress energy-momentum tensor sourced by free-streaming neutrinos. 

Our analysis shows that  the rH framework propagates damped GWs  through the same integro-differential equation of GR but weighted by three new parameters $\delta_{1,2,3}$.  We demostrate that the occurrence of $\delta_3$ can give rise to non-vanishing free-streaming damping effects during the cosmological matter dominated era. 
This behavior could be an important phenomenological signature that is able to highlight the action of modified gravity theories beyond GR. 
Within the context of rH theories, we identify an analytic formula for the multipole orders $\ell_k$ with which free-streaming neutrinos damp the variety of tensor correlation functions of the CMB.

As future generations of  gravitational detectors such as LISA \cite{Audley:2017drz} and DECIGO \cite{Kawamura:2011zz} could get closer to the experimental sensibility to detect cosmological GWs, we deem it of fundamental importance that  the combined impact of the two types of GWs modification studied in this work, i.e. the modification of GR theory and the streaming of free cosmological neutrinos,  be further analyzed and examined to allow for an effective understanding of their data.

\acknowledgments

MS is grateful to his family, A. Pizzol and her family, L. Gobbo and his family, V. Marra, A. Ruzza,  E. Ceron, E. de Lazzari, S. Vidotto, C. Gellussich, A. Possamai, the \textit{Astrofili Veneti} group,  the \textit{Circolo Galileo Galilei} group, the \textit{Vivir Sin Barrio} group,  and the \textit{Aquafit2.0 Master Team 2019}  for their support, inspiration, and motivation. MS also would like to  thank his Professors S. Matarrese, N. Bartolo, and P. Karmakar for their passion and teachings. 
SV wants to thank Vittoria for her beautiful smile.
Some algebraic computations in this article were performed using the Cadabra software \cite{Peeters:2006kp}\cite{Peeters:2007wn}.

\bibliography{bibliography}
\bibliographystyle{JHEP}

\end{document}